\documentclass[aps,prd,reprint,preprintnumbers,groupedaddress,twocolumn,showpacs]{revtex4-1}
\usepackage{graphicx}
\usepackage{bm}

\newcommand{\be}{\begin{equation}}
\newcommand{\ee}{\end{equation}}
\newcommand{\ba}{\begin{array}}
\newcommand{\ea}{\end{array}}
\newcommand{\bqa}{\begin{eqnarray}}
\newcommand{\eqa}{\end{eqnarray}}

\begin{document}

\preprint{ IPPP/10/56, DCPT/10/112, SHEP 10-30, USTC-ICTS-10-13}

\title{Origin of light $0^{+}$ scalar resonances}


\author{Zhi-Yong Zhou}
\email[]{zhouzhy@seu.edu.cn}
\affiliation{Department of Physics, Southeast University, Nanjing 211189,
People's Republic of China}
\affiliation{Institute for Particle Physics Phenomenology,  Durham University, Durham DH1 3LE, United Kingdom}

\author{Zhiguang Xiao}
\email[]{xiaozg@ustc.edu.cn}
\affiliation{ School of Physics and Astronomy,  University of Southampton
  Highfield, Southampton, SO17 1BJ, United Kingdom
}
\affiliation{Interdisciplinary Center for Theoretical Study, University of Science
and Technology of China, Hefei, Anhui 230026, China}

\date{\today}

\begin{abstract}
We demonstrate how most of the light $J^{P}=0^{+}$ spectrum below
$2.0\,\mathrm{GeV}$ and their decays can be consistently described by
the unitarized quark model incorporating the chiral constraints of
Adler zeros and taking $SU(3)$ breaking effects into account. These
resonances appear as  poles in the complex $s$ plane in a unified
picture as $q\bar{q}$ states strongly dressed by hadron loops. Through
the large $N_c$ analysis, these resonances are found to naturally
separate into two kinds: $\sigma, \kappa, f_0(980), a_0(980)$ are
dynamically generated and run away from the real axis as $N_c$
increases, while the others move towards the $q\bar q $ seeds.  In
this picture, the line shape of $a_0(980)$ is produced by a broad pole
below the $K\bar{K}$ threshold, and exhibits characteristics similar to
the $\sigma$ and
$\kappa$.
\end{abstract}


\maketitle

\section{introduction}
The enigmatic spectrum of light $J^{P}=0^{+}$ scalar resonances are of
great interest for its importance in understanding chiral symmetry
breaking and confinement in QCD.  Despite many theoretical efforts,
the current understanding of the microscopic structures of these
resonances is in a well-known unclear situation as summarized in
Particle Data Group~(PDG)~\cite{Amsler:2008zzb}: $q\bar{q}$ models
~\cite{Tornqvist:1995kr,*Tornqvist:1995ay, *Geiger:1992va,
*Boglione:2002vv}, the unitarized meson
model~\cite{vanBeveren:1986ea}, a tetraquark model with and without
$q\bar{q}$ mixing
~\cite{Jaffe:1976ig,*Maiani:2004uc,*Hooft:2008we,Fariborz:2008bd}, the
J\"ulich meson exchange model~\cite{Lohse:1990ew}, the unitarized
$\sigma$ model~\cite{Black:2000qq}, glueball~\cite{Minkowski:1998mf}
or using the inverse amplitude method~(IAM)~\cite{Oller:1998hw}, NJL
model~\cite{Su:2007au} and  lattice simulations~\cite{Alford:2000mm},
and so on. Most of these studies focus on the lowest putative nonet or
explain the lighter and heavier resonances in different ways.  In the
present paper, we show that, all the light scalar spectrum  below
$2.0\,\mathrm{GeV}$ except a glueball candidate can be described (or
even predicted) using just seven parameters in a unified and
consistent picture, that is, $q\bar{q}$ seeds strongly dressed by
hadron loops.  The picture brings more insights on the origin of
the resonances, which are  generated as the poles of the $S$ matrix and have no
one-to-one correspondence with the nonet in the Lagrangian.  At the
weak coupling limit as $N_c$ increases, $\sigma$, $\kappa$, $a_0(980)$,
and $f_0(980)$ move away from the real axis on the complex energy
plane, whereas all the other heavier $J^P=0^+$ resonances move to the
bare seeds. This reveals the differences between the lighter mesons
and heavier states.

We use the unitarized quark
model~(UQM)~\cite{Tornqvist:1995kr,Tornqvist:1995ay} proposed by
T\"ornqvist, which played a pioneering role in the resurrection of
the $\sigma$ meson. The merit of this model is that it naturally respects
the unitarity of the S-matrix but also  incorporates some dynamics at the
same time. Besides, the Adler zeros~\cite{Adler:1965ga}, as the
constraints from chiral symmetry, can also be easily implemented.
Nevertheless, the $\kappa$ resonance was not found in his explicit
analysis of experimental data, and those resonances with higher masses
than $1.5\,\mathrm{GeV}$ are also not covered.  In the present paper,
however, by incorporating the SU(3) breaking effects in the coupling constants and analyzing the poles on the complex plane, we
show that the $\kappa$ resonance can really be found in this picture.
Moreover, most resonances in $I=0$, $I=1/2$, and $I=1$ channels can
find their corresponding poles on the complex plane.

The paper is organized as follows: In Sec.II, we briefly introduce
the basic scheme of UQM and the three nontrivial improvements we make
to this model. Our numerical results are elaborately discussed in
Sec.III. Section IV is devoted to a further study on the
characteristics of these resonances based on the large $N_c$ technique.
Section V summarizes our main results.

\section{The theoretical scheme}
\subsection{Unitarized quark model}
The unitarized quark model begins by assuming that there are  $q\bar
q$ bare bound states generated in QCD and they are coupled with the
pseudoscalar mesons. The main idea is to take into account the hadron
loop dressing effect in the propagators of the bare $q\bar{q}$ states
\cite{Weisskopf:1930au,Tornqvist:1995kr,Boglione:2002vv}. The
bare propagator of a $q\bar q$ bound state is
\bqa
P={1\over m_0^2-s},
\eqa where $m_0$ is the bare mass. For example, $m_0$ is $2\hat{m}$ for
$u\bar{u}$ or $d\bar{d}$, $\hat{m}+\Delta m$ for $u\bar{s}$, and
$2\hat{m}+2\Delta{m}$ for $s\bar{s}$ state, respectively. The vacuum polarization
function, $\Pi(s)$, which represents all the possible two pseudoscalar
meson loops, will contribute to the full propagator as
\bqa
P={1\over m_0^2-s+\Pi(s)}.
\eqa
As an analytic function with a right-hand cut, its real and
imaginary parts are related by a dispersive integral
\be\label{dispRe}
\mathrm{Re}\Pi(s)=\frac{1}{\pi}\mathcal{P}\int_{s_{th}}^\infty\mathrm{d}z{\mathrm{Im}\Pi(z)}/(z-s),
\ee
where
\bqa
\mathrm{Im}\Pi(s)&=&-\sum_i G_i(s)^2\nonumber\\&=&-\sum_i
g_i^2\frac{k_i(s)}{\sqrt{s}}F_i(s)^2\theta(s-s_{th,i}), \eqa where the
general coupling function $G_i(s)$ includes the coupling constants $g_i$'s,
the phase space factor $k_i(s)/\sqrt{s}$, and a Gaussian form factor
$F_i(s)=exp[-k_i^2(s)/2k^2_0]$. $k_i(s)$ is the $i$th channel c.m.
momentum with $k_i(s)=\sqrt{\lambda(s,m_{A_i}^2,m_{B_i}^2)/4s}$ and
the $\theta(s-s_{th,i})$ is a unit step function.

If there exists more than one bare state in the $i\rightarrow j$ channel,
the partial-wave amplitude can be represented in a more general matrix
form:
\bqa
T_{ij}&=&\sum_{\alpha,\beta}G_{i\alpha}P_{\alpha\beta}G^*_{j\beta},\nonumber\\
\{P^{-1}\}_{\alpha\beta}(s)&=&(m_{0,\alpha}^2-s)\delta_{\alpha\beta}+\Pi_{\alpha\beta}(s),\nonumber\\
\mathrm{Im}\Pi_{\alpha\beta}(s)&=&-\sum_iG_{i\alpha }(s)G^*_{\beta i}(s)\nonumber\\
&=&-\sum_i g_{\alpha i}g_{\beta i}\frac{k_i(s)}{\sqrt{s}}F_i^2(s)\theta(s-s_{th,i}),
\eqa
where $\mathrm{Re}\Pi_{\alpha\beta}$ is  determined by a similar
dispersion integral of $\mathrm{Im}\Pi_{\alpha\beta}$ as
Eq.(\ref{dispRe}). The off-diagonal terms of $\Pi_{\alpha\beta}$
produce the mixing between different bare states coupled with the
same intermediate states.

The Adler zeros  are incorporated into the UQM model in a direct and
easily operated phenomenological way~\cite{Tornqvist:1995kr}:
\bqa
G_{\alpha i}(s)G_{\beta i}(s)\rightarrow\gamma_{\alpha i}\gamma_{\beta i}(s-z_{A,i})F_i^2(s)\frac{k_i(s)}{\sqrt{s}}\theta(s-s_{th,i}),\nonumber\\
\eqa
where the $z_{A,i}$'s denote the Adler zeros, and $\gamma_{\beta i}$
are dimensionless coupling constants .

\subsection{The Adler zeros in the Chiral Perturbation Theory}
Normally, the $T$ matrix also contains left-hand cuts. Because the
Adler zero is usually located nearer to the physical threshold than
the left-hand cut, it is natural to expect that it plays a more
important role than the left-hand cut in determining the scattering
amplitudes along the right hand cut.  Since such zeros reflect the
constraints of the chiral symmetry, in the $I=0,1/2$ channels, they are fixed
in our study at the values obtained from the chiral perturbation
theory~(ChPT)~\cite{Gasser:1983yg} rather than left as free
parameters. In the $I=1$ channel, we also fix the Adler zero according to
ChPT but there are some subtleties which will be addressed later.
Being within the convergence radius of the chiral expansion, these
Adler zeros should be reliably determined by ChPT and, hence, is a reasonable
starting point for a phenomenological study.

After partial-wave expansion, one obtains the Adler zero of $I=0$
$\pi\pi$ $S$ wave at about $m^2_\pi/2$ at tree level, and the position is
slightly shifted to $s\simeq 0.38m_\pi^2$ by including the
contribution up to two-loop  $SU(2)$ ChPT\cite{Bijnens:1995yn}.
Similarly, in the $K\pi$ scattering, the $SU(3)$ ChPT to $O(p^2)$ gives
S-wave amplitude of $I=1/2$ as
\bqa
T^{l=0}_{I=1/2}(s)=\frac{-3{\left( {m_K}^2 -
         {m_\pi}^2 \right) }^2 -
    2\left( {m_K}^2 +
       {m_\pi}^2 \right) s + 5s^2}{128\pi
    {f_K}{f_\pi} s},\nonumber\\
\eqa
which has two zeros located at $\frac{1}{5}({m_K}^2 + {m_\pi}^2 \pm
    2{\sqrt{4{m_K}^4 -
         7{m_K}^2\,{m_\pi}^2 +
4{m_\pi}^4}})$
in the unphysical region. One is on the negative real axis
inside the circular cut, and the other is on the positive axis between
the circular cut and the $K\pi$ threshold.  With the $O(p^4)$ contribution,
the left zero on the negative axis moves to the complex $s$ plane at about
$s=0.003\pm 0.005i\mathrm{GeV}^2$ and the right one on the positive axis
will also be slightly shifted (at about $s=0.233\mathrm{GeV}^2$). 
 As for the $\pi\eta$ scattering, at leading order
$O(p^2)$, ChPT recovers the current-algebra
result\cite{Bernard:1991xb}:
\bqa
T^{(2)}_{\pi\eta}(s,t,u)=\frac{m_\pi^2}{3f_\pi^2},
\eqa
which contains
no zero on the real $s$ axis after the partial-wave projection. Including the higher
order contribution is not helpful to obtain a real-valued Adler zero.
A pair of zeros, at about $0.078\pm i0.178\,\mathrm{GeV^2}$, can be
found when the $O(p^4)$ terms are taken into account using the
low-energy constants from Ref.\cite{Gasser:1984gg}. Since there is no
accurate data of the $\pi\eta$ scattering, we could choose a zero point on the real
axis to simulate them as mentioned later.

\subsection{The scalar-pseudoscalar-pseudoscalar coupling}
The coupling of pseudoscalar-pseudoscalar to the $0^{+}$ states could be
described by effective interaction terms in the Lagrangian: $\mathcal
{L}_{SPP}=\alpha Tr[SPP]+\beta Tr[S]Tr[PP]+\gamma Tr[S]Tr[P]Tr[P]$. The first
term has an $SU(3)$-symmetric quarkonium coupling as used in many
phenomenological models.  We also include the Okubo-Zweig-Iizuka violation terms in the last two
terms in a general way. Moreover, the decay of the quarkonium into a pair of
mesons $Q\bar{Q}\rightarrow M(Q\bar{q}_i)M(q_i\bar{Q})$ involves the creation
of a $q_i\bar{q}_i$ pair from the vacuum. The ratio of the creation
rates of $s\bar{s}$ and $u\bar{u}$ or $d\bar{d}$ from the vacuum is
usually defined as
$\rho={\langle0|V|s\bar{s}\rangle}/{\langle0|V|u\bar{u}\rangle}$,
representing the breaking of $SU(3)$ symmetry~\cite{Amsler:1995td}.
$SU(3)$ breaking effects have proved to be important, so we allow for
these in our version of the UQM. To be explicit about our description
of the coupling of quarkonium to mesons, we express the scalar and
pseudoscalar $3\times 3$ flavor matrices in $q\bar{q}$ configurations
as
\bqa
 S&=&\left(
  \begin{array}{ccc}
    {1\over \sqrt{2}}a^0+{1\over \sqrt{2}}f_n & a^+ & \kappa^+ \\
    a^- & {-1\over \sqrt{2}}a^0+{1\over \sqrt{2}}f_n & \kappa^0 \\
    \kappa^- & \bar{\kappa}^0 & f_s \\
  \end{array}
\right),
\\ P&=&\left(
  \begin{array}{ccc}
    {\sqrt{3}\pi^0+\eta_8\over \sqrt{6}} & \pi^+ & K^+ \\
    \pi^- & {-\sqrt{3}\pi^0+\eta_8\over \sqrt{6}}& K^0 \\
    K^- & \bar{K}^0 & -{\sqrt{2\over 3}}\eta_8 \\
  \end{array}
\right)+{1\over \sqrt{3}}\eta_1,
\eqa
where $f_n=n\bar{n}\equiv (u\bar{u}+d\bar{d})/\sqrt{2}$ and
$f_s=s\bar{s}$.
The physical states, $\eta$ and $\eta'$, are conventionally defined as
$
\eta=\mathrm{cos}\phi|n\bar{n}\rangle-\mathrm{sin}\phi|s\bar{s}\rangle
$,
$\eta'=\mathrm{sin}\phi|n\bar{n}\rangle+\mathrm{cos}\phi|s\bar{s}\rangle,
$ with $\phi=\mathrm{tan}^{-1}\sqrt{2}+\theta_{P},$ where
$\theta_{P}=-11.5^\circ$ being the pseudoscalar octet-singlet mixing
angle~\cite{Amsler:2008zzb}. Thus, by standard derivation, a general form of
effective coupling constants between the scalar quarkonium and pseudoscalar pair is
obtained, as shown in Table~\ref{su3c}.

\begin{table}
\caption{\label{su3c}The effective scalar quarkonium coupling to pseudoscalar mesons up to a global
constant.
}
\begin{center}
\begin{tabular}{|c|c|c|}
  \hline
  I &   & Coupling coefficient \\
  \hline
  0($n\bar{n}$) & $\pi\pi$ & $-\sqrt{3}\alpha-2\sqrt{3}\beta$ \\
  & $K\bar{K}$ & $-\rho \alpha-4\beta$ \\
   & $\eta\eta$ & $\mathrm{cos}\phi^2 \alpha+2\beta+2(1+\mathrm{cos}\phi^2-\sqrt{2}\mathrm{sin}2\phi)\gamma$ \\
   & $\eta\eta'$ & $\frac{\mathrm{sin}2\phi}{\sqrt{2}}\alpha+(4\mathrm{cos}2\phi+\sqrt{2}\mathrm{sin}2\phi)\gamma$ \\
   & $\eta'\eta'$ & $\mathrm{sin}\phi^2\alpha+2\beta+2(1+\mathrm{sin}\phi^2+\sqrt{2}\mathrm{sin}2\phi)\gamma$ \\
  \hline
  0($s\bar{s}$) & $\pi\pi$ & $-\sqrt{6}\beta$ \\
  & $K\bar{K}$ & $-\sqrt{2}\alpha-2\sqrt{2}\beta$ \\
   & $\eta\eta$ & $\sqrt{2}(\rho \mathrm{sin}\phi^2 \alpha+\beta+(1+\mathrm{cos}\phi^2-\sqrt{2}\mathrm{sin}2\phi)\gamma)$ \\
   & $\eta\eta'$ & $- \rho\mathrm{sin}2\phi \alpha+2\sqrt{2}\mathrm{cos}2\phi \gamma+\mathrm{sin}2\phi \gamma$ \\
   & $\eta'\eta'$ & $\sqrt{2}(\rho\mathrm{cos}\phi^2 \alpha+\beta+(1+\mathrm{sin}\phi^2+\sqrt{2}\mathrm{sin}2\phi)\gamma)$ \\
  \hline
   1 & $\eta\pi$ & $\sqrt{2}\mathrm{cos}\phi \alpha$ \\
   & $K\bar{K}$ & $-\rho \alpha$ \\
   & $\eta'\pi$ & $\sqrt{2}\mathrm{sin}\phi \alpha$ \\
  \hline
  1/2 & $K\pi$ & $-\sqrt{3/2} \alpha$ \\
   & $K\eta$ & $(\frac{\mathrm{cos}\phi}{\sqrt{2}}-\rho\mathrm{sin}\phi)\alpha$ \\
   & $K\eta'$ & $(\frac{\mathrm{sin}\phi}{\sqrt{2}}+\rho\mathrm{cos}\phi)\alpha$ \\
  \hline
  \end{tabular}
\end{center}
\end{table}

\subsection{The analyticity of $S$ matrix}
By definition, a resonance is specified as a pole of the $S$ matrix
analytically continued to the complex $s$ plane. Extracting the poles
of the partial-wave amplitude of $i\rightarrow j$ process described in
the UQM is actually to find the zeros of the determinant of the
inverse propagator. Sometimes, it can be obtained in some other
equivalent way. For example, in a two-channel occasion the
analytically continued $S$ matrices on different Riemann sheets could be
written down using those on the first sheet\cite{Xiao:2001pt}:
\bqa
S^{II}&=&\left(
         \begin{array}{cc}
           \frac{1}{S_{11}} & \frac{iS_{12}}{S_{11}} \\
            \frac{iS_{12}}{S_{11}} &  \frac{detS}{S_{11}} \\
         \end{array}
       \right),
S^{III}=\left(
         \begin{array}{cc}
           \frac{S_{22}}{detS} & \frac{-S_{12}}{detS} \\
            \frac{-S_{12}}{detS} &  \frac{S_{11}}{detS} \\
         \end{array}
       \right),\nonumber\\
       S^{IV}&=&\left(
         \begin{array}{cc}
           \frac{detS}{S_{22}} & -\frac{iS_{12}}{S_{22}} \\
            -\frac{iS_{12}}{S_{22}} &  \frac{1}{S_{22}} \\
         \end{array}
       \right),
\eqa
which implies that a pole on the second Riemann sheet is just located
at a zero point of $S_{11}$ on the first sheet, a third-sheet pole
at a zero point of $detS$, and a fourth-sheet pole at a zero of
$S_{22}$, respectively.
\begin{figure}[t]%
\begin{center}%
\hspace{-2cm}\includegraphics[height=30mm]{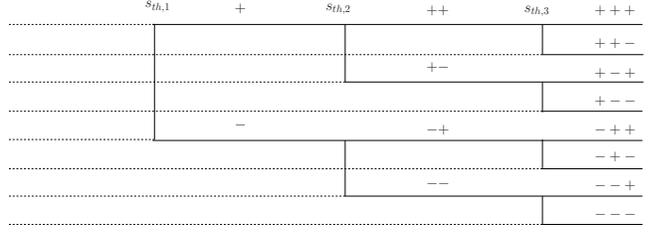}
\caption{\label{multicut} The right-hand cuts and their signatures in a three-channel case.}
\end{center}%
\end{figure}%

In the literature it is common to define a Breit-Wigner mass of a
resonance as the solution of $m_{BW}^2=m_B^2+{ Re}\Pi(m_{BW}^2)$.
This is a good approximation for narrow resonances and is commonly
used in experimental analysis.  However, if the propagator is strongly
dressed by hadron loops where $Im\Pi(s)$ is large, the mass and width
are no longer suitably determined by the Breit-Wigner form, but should
only be defined by the pole position of the $S$ matrix, i.e., the
solution of $m_B^2-s_p+\Pi(s_p)=0$ with $s_p=(M-i\Gamma/2)^2$.
Because of the analyticity of the $S$ matrix, the determinant of
inverse propagator vanishes only on unphysical Riemann sheets.

The general character of the poles on different Riemann sheets has
been discussed widely in the literature, (see, for
example,~\cite{newton:1961,Eden:1964zz}). Every physical cut will double
the Riemann sheets in the analytical continuation, so there are $2^n$
Riemann sheets in a process with $n$ coupled channels, as shown in
Fig.\ref{multicut}. The physical sheet is defined as the sheet where
all the c.m. momenta are positive on the physical cuts, denoted as
$(+++\cdots +)$ signature.  In the same fashion, the $(n+1)^{th}$
sheet ($n\leq N$), attached to the physical sheet between $s_{th,n}$
and $s_{th,n+1}$ along the physical cut, is denoted by
$(-\cdots-++\cdots)$ with the $n$ consecutive ``$-$" signs before the other
``$+$" signs.  A resonance is represented by a pair of conjugate poles
on the Riemann sheet, as required by the real analyticity. The micro-causality
tells us the first Riemann sheet is free of complex-valued poles, and
the resonances are represented by those poles on unphysical sheets.
The resonance behavior is only significantly influenced by those
nearby poles, and that is why only those closest poles to the
experiment region could be extracted from the experiment data in
a phenomenological study.  Those poles on the other  $n>N$ sheets, which are
reached indirectly, make less contribution and are thus harder to
determine.

\section{Numerical analysis}

\begin{figure}[t]
\hspace{-1cm}\includegraphics[height=30mm]{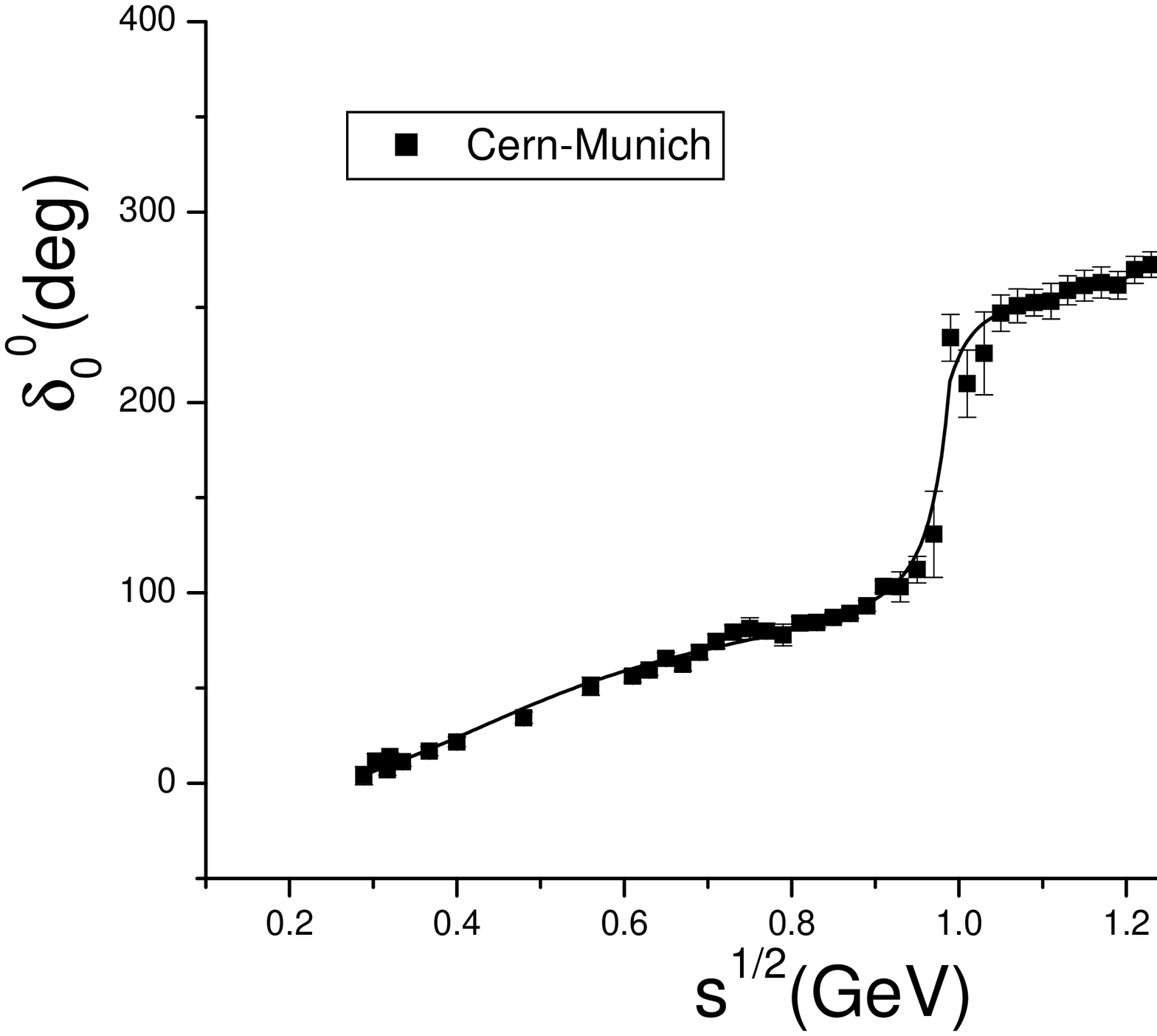}
\hspace{1cm}\includegraphics[height=30mm]{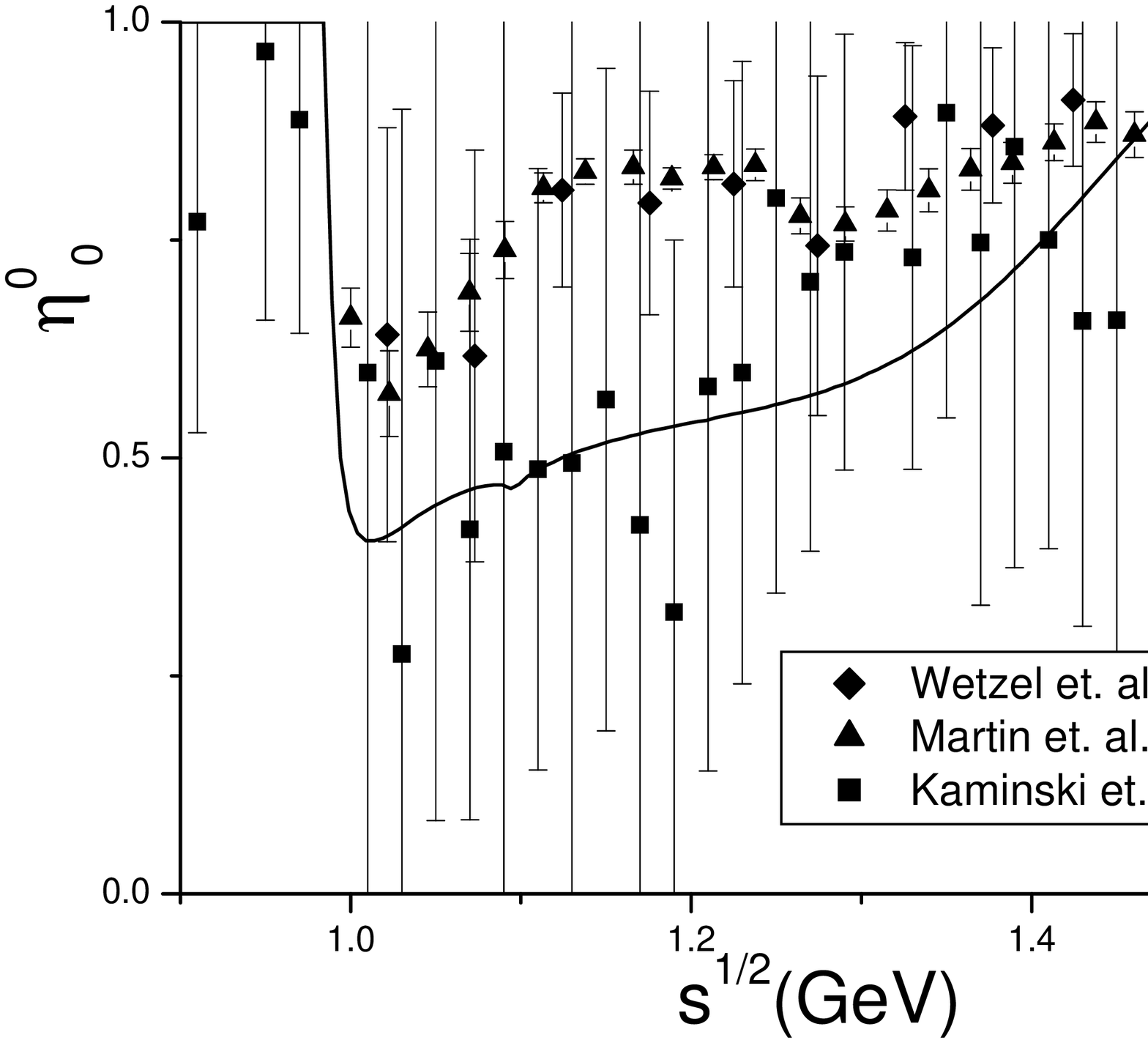}\\
\hspace{-1cm}\includegraphics[height=30mm]{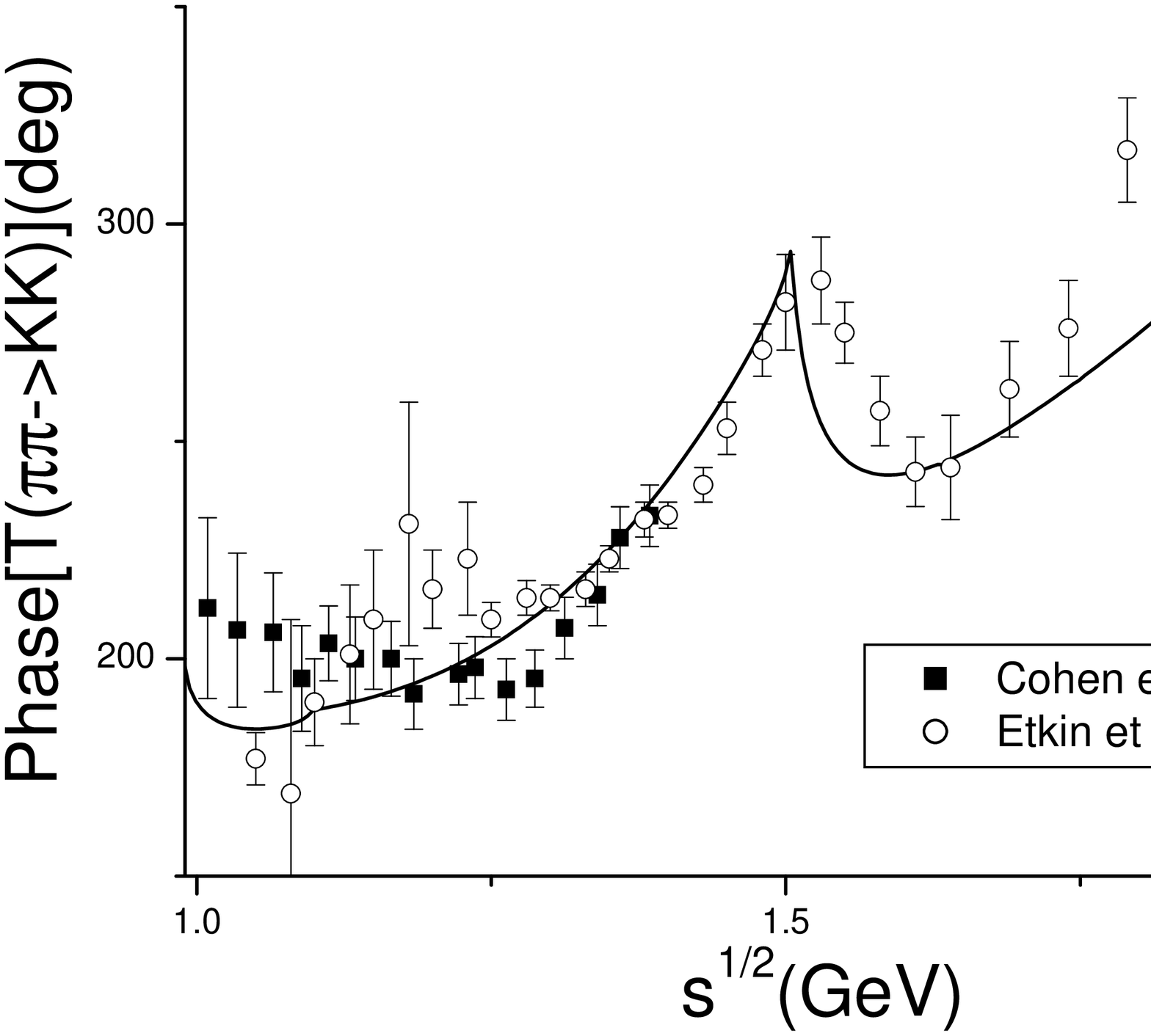}
\hspace{1cm}\includegraphics[height=30mm]{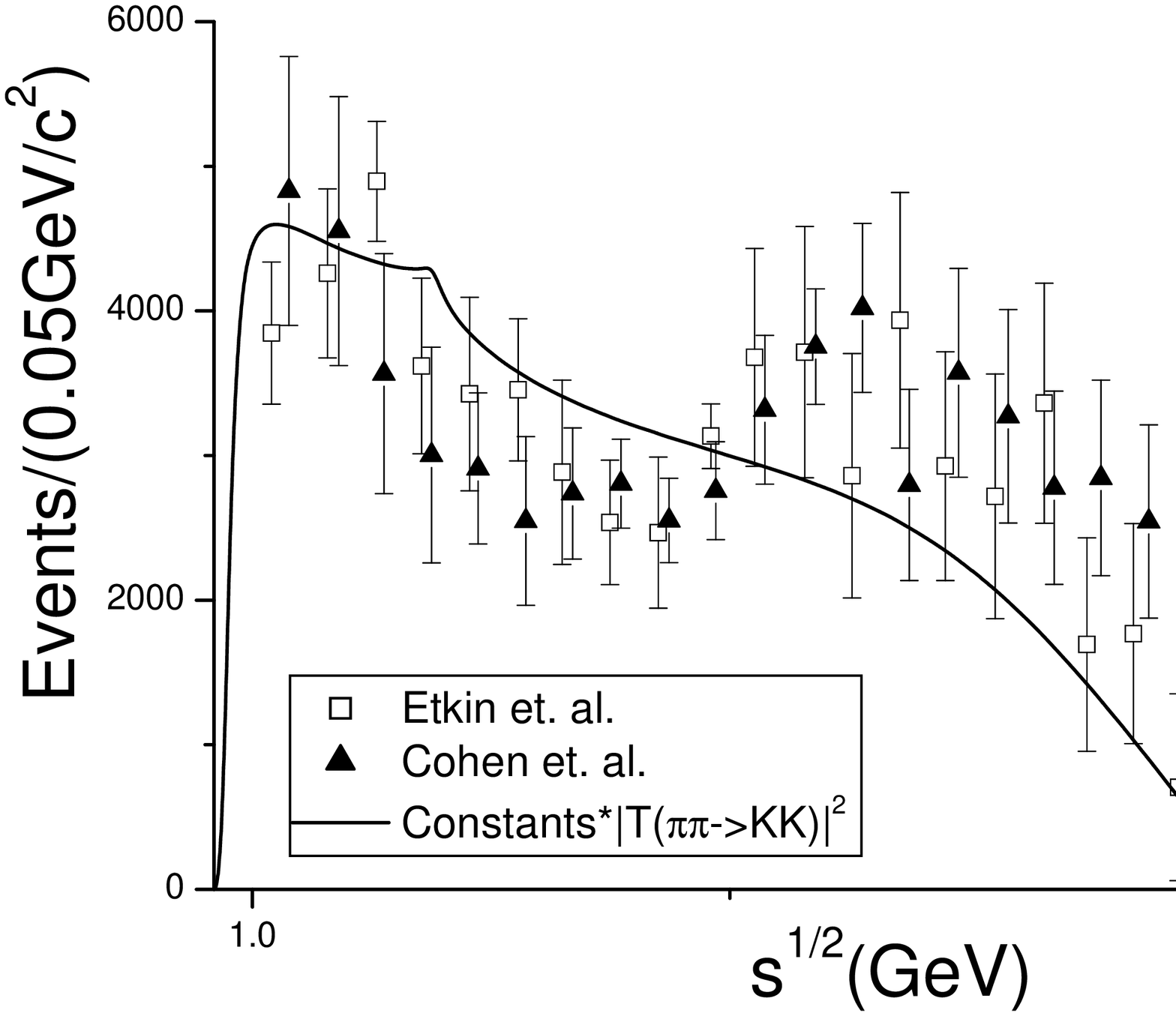}\\
\hspace{-1cm}\includegraphics[height=30mm]{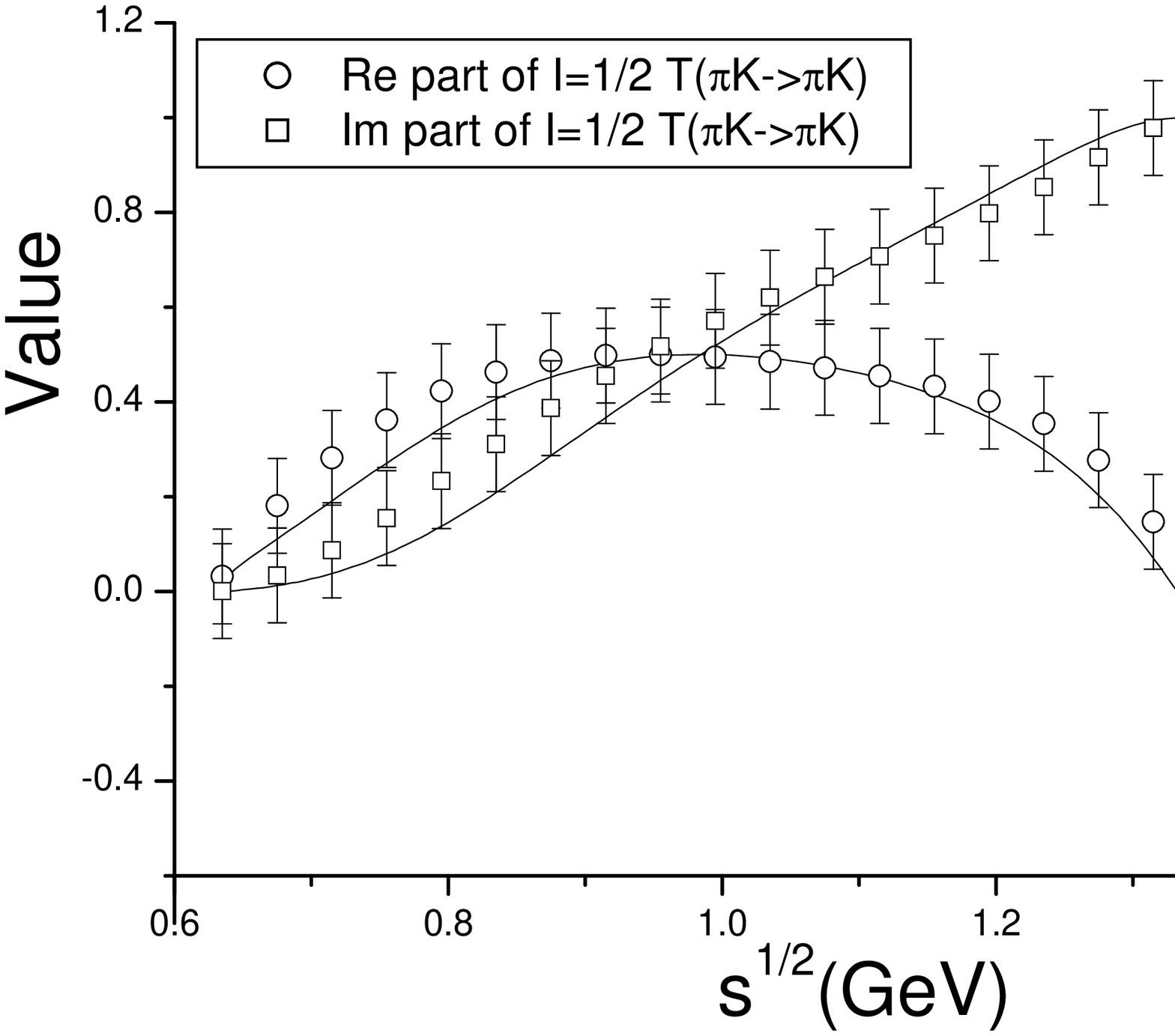}
\hspace{1cm}\includegraphics[height=30mm]{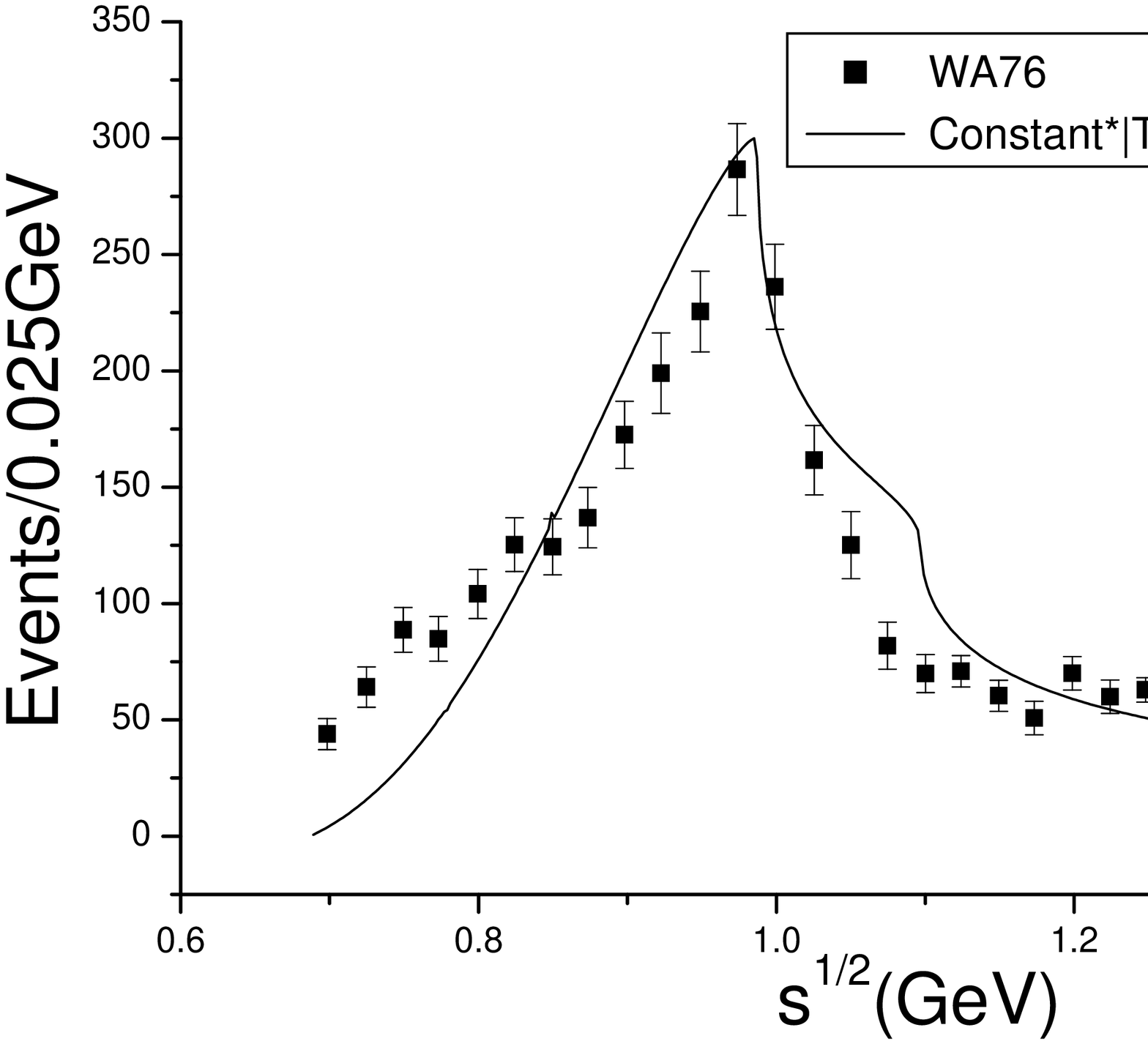}
\caption{\label{expcmp}  The left three figures show the fit quality. The
right three ones are predictions. Compared data are
from~\cite{Cohen:1980cq,*Etkin:1981sg,Wetzel:1976gw,*Martin:1979gm,*Kaminski:1996da,*Armstrong:1991rg}.}
\end{figure}

\begin{figure}[t]
\hspace{-1cm}\includegraphics[height=30mm]{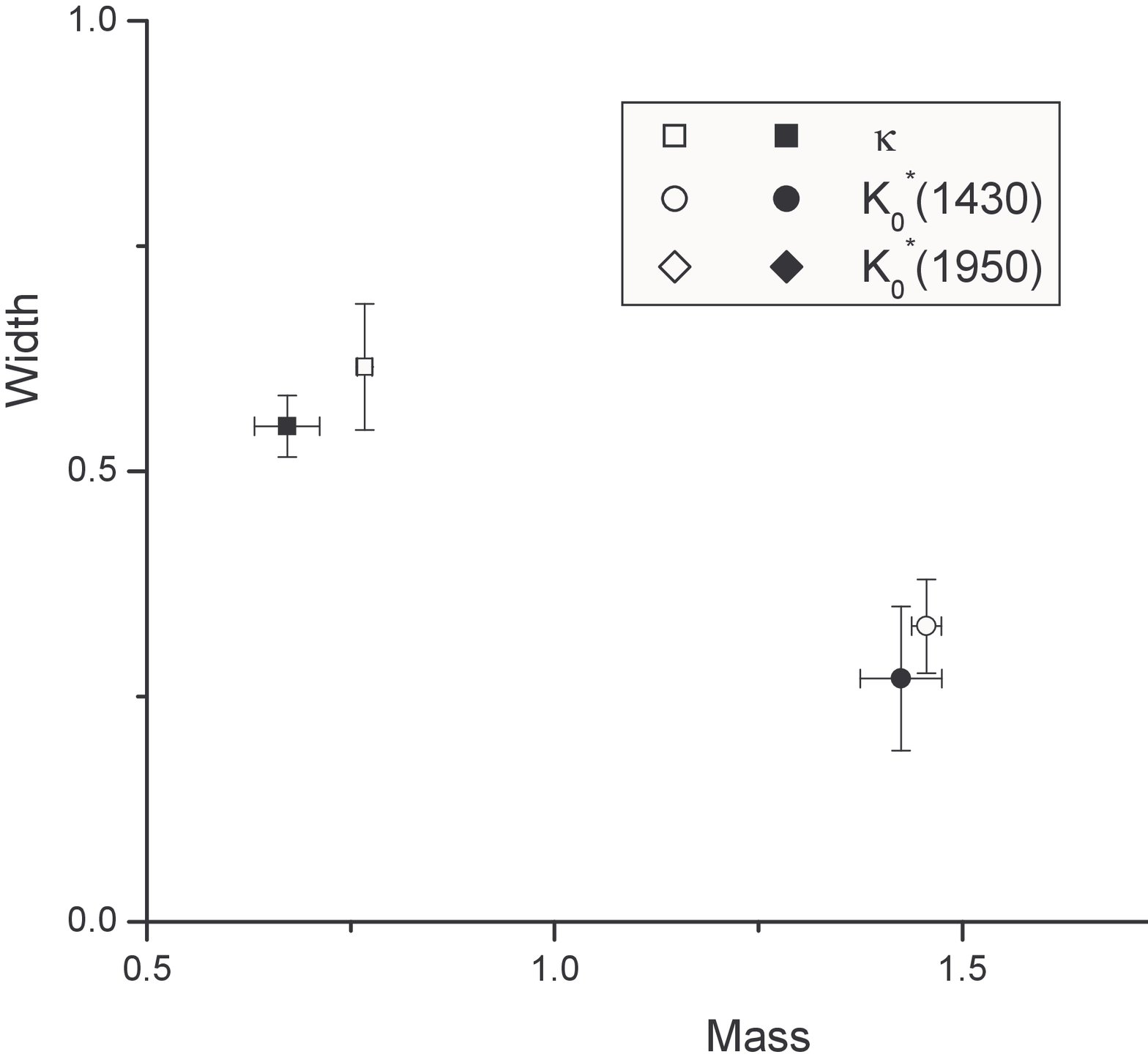}
\hspace{1cm}\includegraphics[height=30mm]{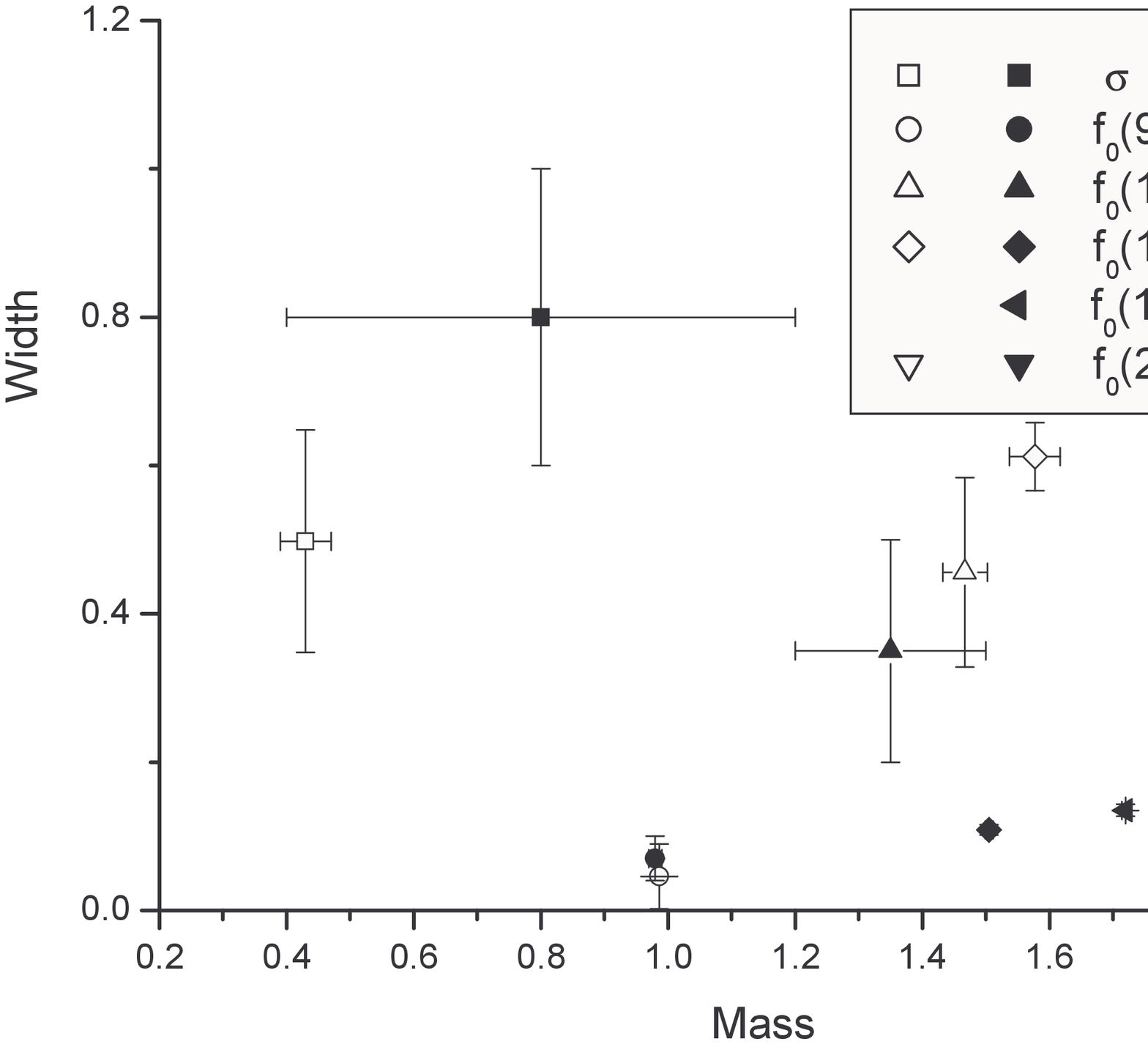}\\
\includegraphics[height=30mm]{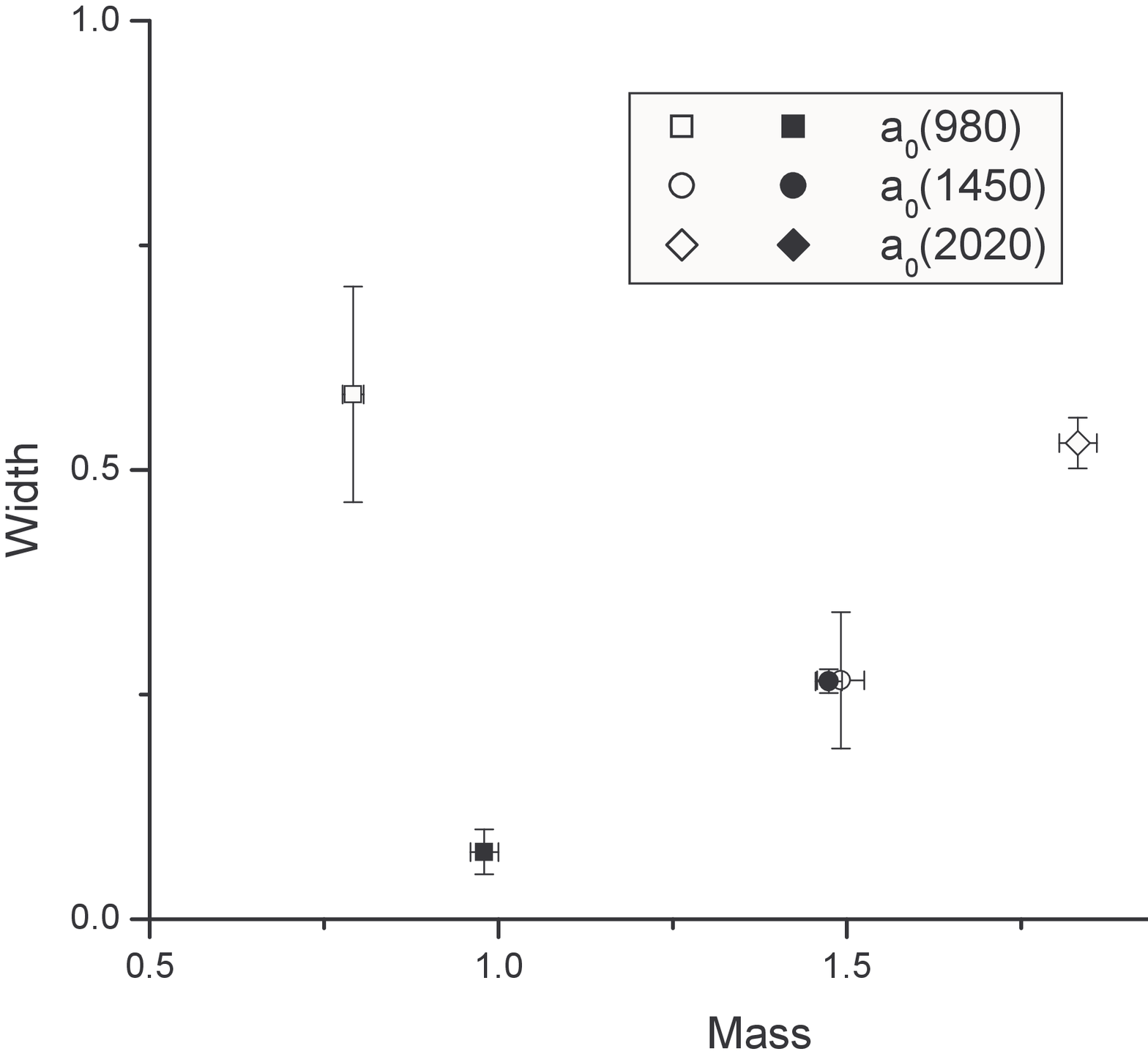}
\caption{\label{compile}
The filled symbols represent the resonances' masses and widths from
the PDG table, and the empty ones represent the pole masses and widths we
obtained. The reason of the discrepancy between the value sets has
been discussed in the text.}
\end{figure}

Now, we apply the partial-wave formulation~\cite{Tornqvist:1995kr}
with our new ingredients to study $I=1/2$ $K\pi$, $I=0$ $\pi\pi$, and
$I=1$ $\pi\eta$ S-wave scattering.  The main purpose of this paper is
not to make an exhaustive fit, so the only data used in the combined
fit are: (1)~the $I=1/2$ S-wave $K\pi$ scattering
amplitude~\cite{Aston:1987ir,Pennington:2007se}, (2)~the phase shift
of $I=0$ S-wave $\pi\pi$ scattering~\cite{Ochs:73,*Grayer:1974cr}, and
(3)~the phase shift of $\pi\pi\rightarrow
K\bar{K}$~\cite{Cohen:1980cq,*Etkin:1981sg} below $1.5\,\mathrm{GeV}$.
The fit quality is good and $\chi^2/d.o.f. \simeq0.8$. The central
values and the statistical errors of the seven parameters are listed in
Table~\ref{para}. The good agreement between our theoretical results
and experimental data can be seen in Fig.~\ref{expcmp}, even though
some of these data have not been used in the fit.  The parameter
values in Table~\ref{para} are all in realistic ranges. The bare
masses of $q\bar{q}$ states are slightly larger, but not in conflict
with NJL modelings~\cite{Su:2007au}.  The $SU(3)$ breaking effect
parameter is also consistent with the value in the
literature~\cite{Amsler:1995td}.  A general comparison of the masses
and widths of the resonances from our results and the values from the
PDG table is presented in Fig.~\ref{compile} which will be discussed
in detail.
\begin{table}[t]
\caption{\label{para}The results of the fit parameters with a Gaussian form factor. $m_0$, $m_0+m_s$, and $m_0+2 m_s$
 are the  bare masses of $n\bar{n}$, $n\bar{s}$, and $s\bar{s}$ respectively.  }
\begin{center}
\begin{tabular}{|c|c|}
  \hline
  $\alpha$ :  $\beta$ :  $\gamma$   &  $1.493_{\pm0.051}$ : ($-0.149_{\pm 0.025}$) : $0.319_{\pm 0.021}$\\
  \hline
  $\rho$ & $0.704_{\pm0.054}$\\
  \hline
   $k_0$(GeV)&  $0.505_{\pm0.009}$ \\
  \hline
   $m_0$(GeV) &$1.443_{\pm0.020}$\\
  \hline
   $m_s$(GeV)&$0.046_{\pm0.012}$ \\
  \hline
  \end{tabular}
\end{center}
\end{table}

The $I=1/2$ $K\pi$ S-wave scattering provides an ideal illustration and
the best testing ground because of its large threshold spans and clean
experimental information.  With the parameters of Table~\ref{para},
there is  only one solution to $s=(m_0+m_s)^2 +Re\Pi(s)$ for real $s$
at about $s^{1/2}=1.33\,\mathrm{GeV}$, which is the Breit-Wigner-like
mass mentioned previously. However, there are three poles of the
$S$ matrix found near the physical region  at
\bqa\label{pikpoles}
\sqrt{s^{II}}&=&0.767_{\pm0.009}-i 0.308_{\pm0.035}  ,\nonumber\\ \sqrt{s^{III}}&=&1.456_{\pm0.018}-i 0.164_{\pm0.026} , \nonumber\\
  \sqrt{s^{IV}}&=&1.890_{\pm0.029}-i 0.296_{\pm0.014} ,
\eqa
where the superscript denotes the number of the sheet and the units
are in $\,\mathrm{GeV}$.  Simply comparing these poles with the tables
of Particle Data Group~\cite{Amsler:2008zzb}, a good agreement in
quality is instantly found (see Fig.~\ref{compile}). The
lowest second-sheet pole is just the $\kappa$ resonance and consistent
with the values determined by those model-independent
methods~\cite{Zheng:2003rw,*Zhou:2006wm,*DescotesGenon:2006uk}. The
third-sheet pole corresponds to the $K_0^*(1430)$ and a second-sheet
``shadow" pole (due to the weak coupling constant of the $K\eta$ channel in
our result and also found in~\cite{Eden:1964zz}) is also found at
almost the same location.  Although the fit is only carried out with
the data below $1.5\,\mathrm{GeV}$, a fourth-sheet pole is predicted
by the $SU(3)$ couplings and unitarity constraints, which corresponds
to the higher $K_0^*(1950)$ resonance.  The width of $K^*_0(1950)$ is
larger than its PDG value, but is qualitatively acceptable compared with
the average
value calculated from both solutions A and B of the original data
analysis~\cite{Aston:1987ir,Anisovich:1997qp,*Jamin:2000wn}.

The poles in $I=0$ $\pi\pi$ $S$ wave are inevitably more complicated
than those in the $K\pi$ $S$ wave because of the mixing of $n\bar{n}$
and $s\bar{s}$ states. We find
\bqa
\sqrt{s^{II}}&=&0.430_{\pm0.040}-i 0.249_{\pm0.075} ,\nonumber\\
\sqrt{s^{II}}&=&0.986_{\pm0.015}-i 0.023_{\pm0.022} ,  \nonumber\\
  \sqrt{s^{IV}}&=&1.467_{\pm0.035}-i 0.228_{\pm0.064} ,\nonumber\\
\sqrt{s^{V}}&=&1.577_{\pm0.040}-i 0.306_{\pm0.023} ,
\nonumber\\
\sqrt{s^{VI}}&=&1.935_{\pm0.028}-i 0.289_{\pm0.013} ,\nonumber\\
\sqrt{s^{VI}}&=&2.444_{\pm0.032}-i 0.242_{\pm0.013} .
\eqa
All these poles could be assigned to those light resonances of
$I^G(J^{PC})=0^+(0^{++})$ listed in the PDG table, except for the
$f_0(1710)$.  The position of the $\sigma$ pole is in agreement with the
results of model-independent
analysis~\cite{Zhou:2004ms,*Caprini:2005zr}, while the $f_0(980)$ is a
narrow pole below the $K\bar{K}$ threshold.  The $f_0(1370)$ is a fourth-sheet
pole and its position is within the uncertainty of the PDG value. The
pole mass is consistent with that preferred by the Belle Collaboration
from $\gamma\gamma\rightarrow\pi^0\pi^0$, at
1.47GeV~\cite{Uehara:2008pf}.  Here, the resonance shape of
$f_0(1500)$ is generated by the fifth-sheet pole and $\eta\eta'$
threshold together, as found in Ref.~\cite{Albaladejo:2008qa}.
This may be the reason why the width of the pole is much wider than
the PDG value.  The $4\pi$
and $\rho\rho$ thresholds turn out to be increasingly important beyond
about $1.2\,\mathrm{GeV}$. However, the inclusion of such thresholds
requires the $SVV$ and $VPP$ interactions to be taken into account. This
would introduce many new parameters and this case is beyond the scope of this
paper. Not incorporating the $SVV$ interaction might explain why there is
no $f_0(1710)$ pole in this picture.  The other possibility is that
the main ingredient of $f_0(1710)$ could be the lowest scalar
glueball, as preferred by recent quenched Lattice
calculations~\cite{Lee:1999kv,*Liu:2000ce,*Chen:2005mg}. This would
add a narrow resonance structure with its own hadron
cloud~\cite{Boglione:1997aw}. The two sixth-sheet poles are assigned
to the $f_0(2020)$ and the $f_0(2330)$, respectively.

As for the $I=1$ $\pi\eta$ scattering, owing to the poorness of data, this
channel is not included in the fit and so the plot in the lower right
corner of Fig.~1 is wholly a prediction. As we have mentioned
previously, no real-value Adler zero is found up to $O(p^4)$ ChPT
amplitude in this partial wave.  We simply set
$z_A=0.078\,\mathrm{GeV}^2$ in the calculation, which is close to the
complex zeros. It is not difficult to exhibit the $a_0(980)$ line
shape below the $K\bar{K}$ threshold, as shown in Fig.~\ref{expcmp}.  The
pole positions are not sensitive to small deviations from the value of Adler zero we choose:
\bqa
\sqrt{s^{II}}&=&0.792_{\pm0.015}-i 0.292_{\pm0.060}  ,\nonumber\\
\sqrt{s^{III}}&=&1.491_{\pm0.034}-i 0.133_{\pm0.038}  , \nonumber\\
\sqrt{s^{IV}}&=&1.831_{\pm0.027}-i 0.265_{\pm0.014} .
\eqa
The second-sheet pole below the $K\bar{K}$ threshold is broad, but still
produces the $a_0(980)$ line shape combined with the threshold effect,
as proposed by Flatt$\acute{\mathrm{e}}$~\cite{Flatte:1976xu}. This
effect was also found by studying the $\pi\eta$ amplitude in the J\"ulich
model~\cite{Janssen:1994wn} and implied in the unitarized $\sigma$
model~\cite{Black:2000qq}.  The third-sheet pole could represent
the $a_0(1450)$, although it plays a minor role in our picture.  There is
also a higher $a_0(1830)$ predicted, which has not been widely
observed in experiments, but might be related to the $a_0(2020)$ seen
by the Crystal Barrel Collaboration~\cite{Anisovich:1999jv}.

\begin{table}[t]
\caption{\label{para2}The results of the fit parameters with another  form factor.}
\begin{center}
\begin{tabular}{|c|c|}
  \hline
  $\alpha$ :  $\beta$ :  $\gamma$   &  $1.889_{\pm0.024}$ : ($-0.280_{\pm 0.015}$) : $0.269_{\pm 0.011}$\\
  \hline
  $\rho$ & $0.821_{\pm0.027}$\\
   \hline
   $m_0$(GeV) &$1.763_{\pm0.038}$\\
  \hline
   $m_s$(GeV)&$0.215_{\pm0.037}$ \\
  \hline
  \end{tabular}
\end{center}
\end{table}

The use of a Gaussian form factor is the most drastic assumption we
have made.  This has been widely used in experimental analyses and in
many other models. However, an exact representation of the form
factor, satisfying unitarity and analyticity and easily applied in
phenomenological studies, has not been found.  To test the stability
of our results, we also use a different form factor,
$[(M^2+s_{th,i})/(M^2+s)]^2$ ($M$ is the mass of the resonance)
proposed in ~\cite{Ravndal:1972pn}. The latter form factor is not
better than the Gaussian form factor since it suffers from a
spacelike pole, but it might provide a reasonable qualitative
cross-check, especially for the $N_c$ analysis we addressed later.
Incorporating the latter form factor  removes one parameter, $k_0$,
so the fit quality is worse and $\chi^2/d.o.f.\simeq 1.21$.
The central values of the other six parameters listed in Table \ref{para2} are different from those in Table \ref{para}.   Nevertheless,  the poles below
$1.5\,\mathrm{GeV}$, as shown in the following, coincide in position
with those found using the Gaussian form factor:
\bqa\label{pikpoles2}
\sqrt{s^{II}}&=&0.745_{\pm0.007}-i 0.301_{\pm0.023}  ,\nonumber\\ \sqrt{s^{III}}&=&1.547_{\pm0.021}-i 0.148_{\pm0.026} ,
\eqa
for the $I=1/2$ channel,
\bqa
\sqrt{s^{II}}&=&0.396_{\pm0.019}-i 0.244_{\pm0.046} ,\nonumber\\  \sqrt{s^{II}}&=&0.984_{\pm0.023}-i 0.028_{\pm0.026} ,  \nonumber\\
  \sqrt{s^{IV}}&=&1.455_{\pm0.037}-i 0.304_{\pm0.040} ,
\eqa
for the $I=0$ channel and
\bqa
\sqrt{s^{II}}&=&0.700_{\pm0.060}-i 0.265_{\pm0.031},
\eqa
for the $I=1$ channel. The corresponding poles beyond $1.5$ GeV still
exist, but since we only fit the data below $1.5$ GeV and have not included
the $I=1$ data, they move farther away from the previous
values. $A$ $posteriori$, this means the second kind of form factor may
not be a good choice.

\section{Large $N_c$ analysis of the pole trajectories}
\begin{figure}[t]
\hspace{-1.cm}\includegraphics[height=30mm]{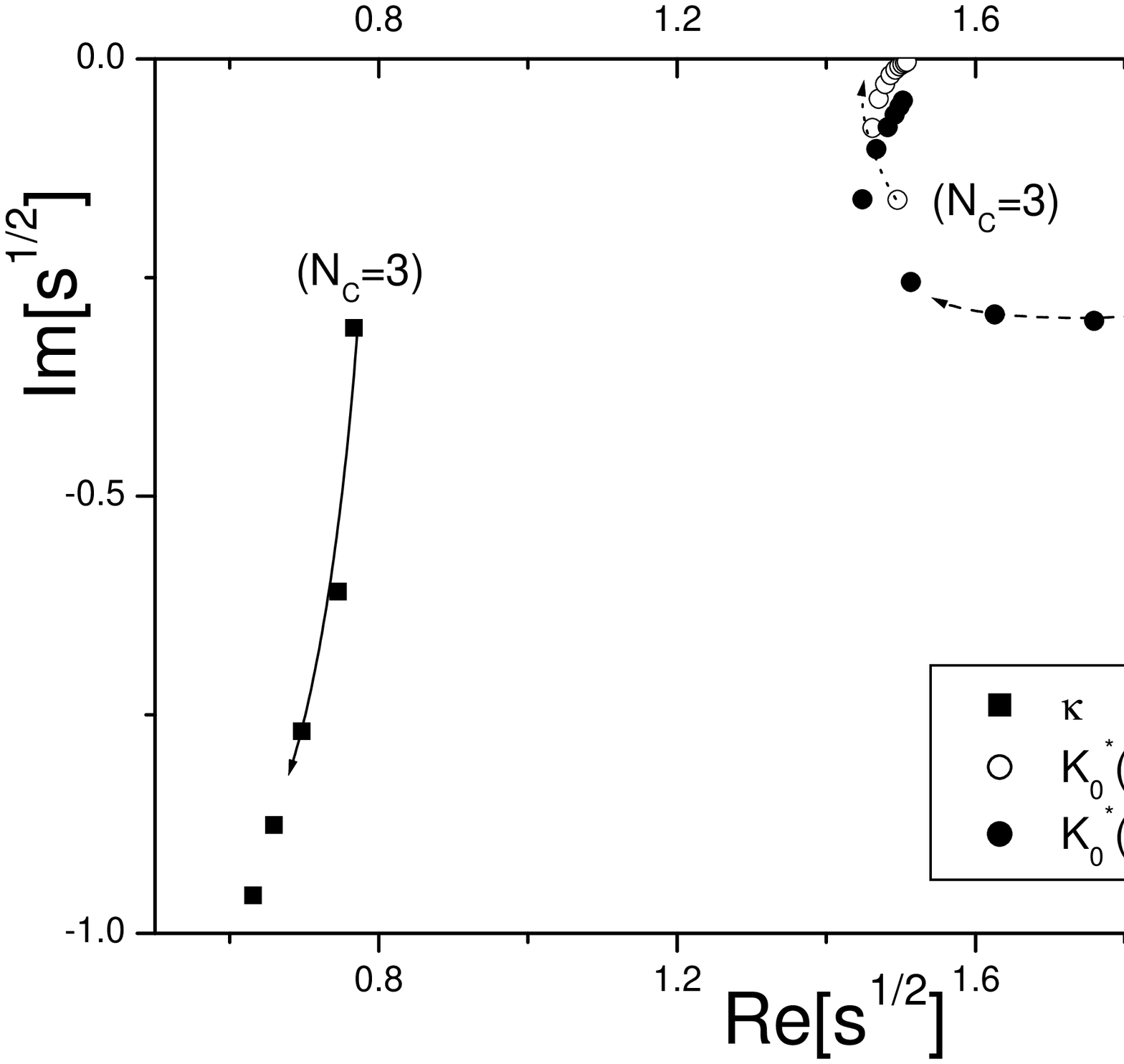}
\hspace{1cm}\includegraphics[height=30mm]{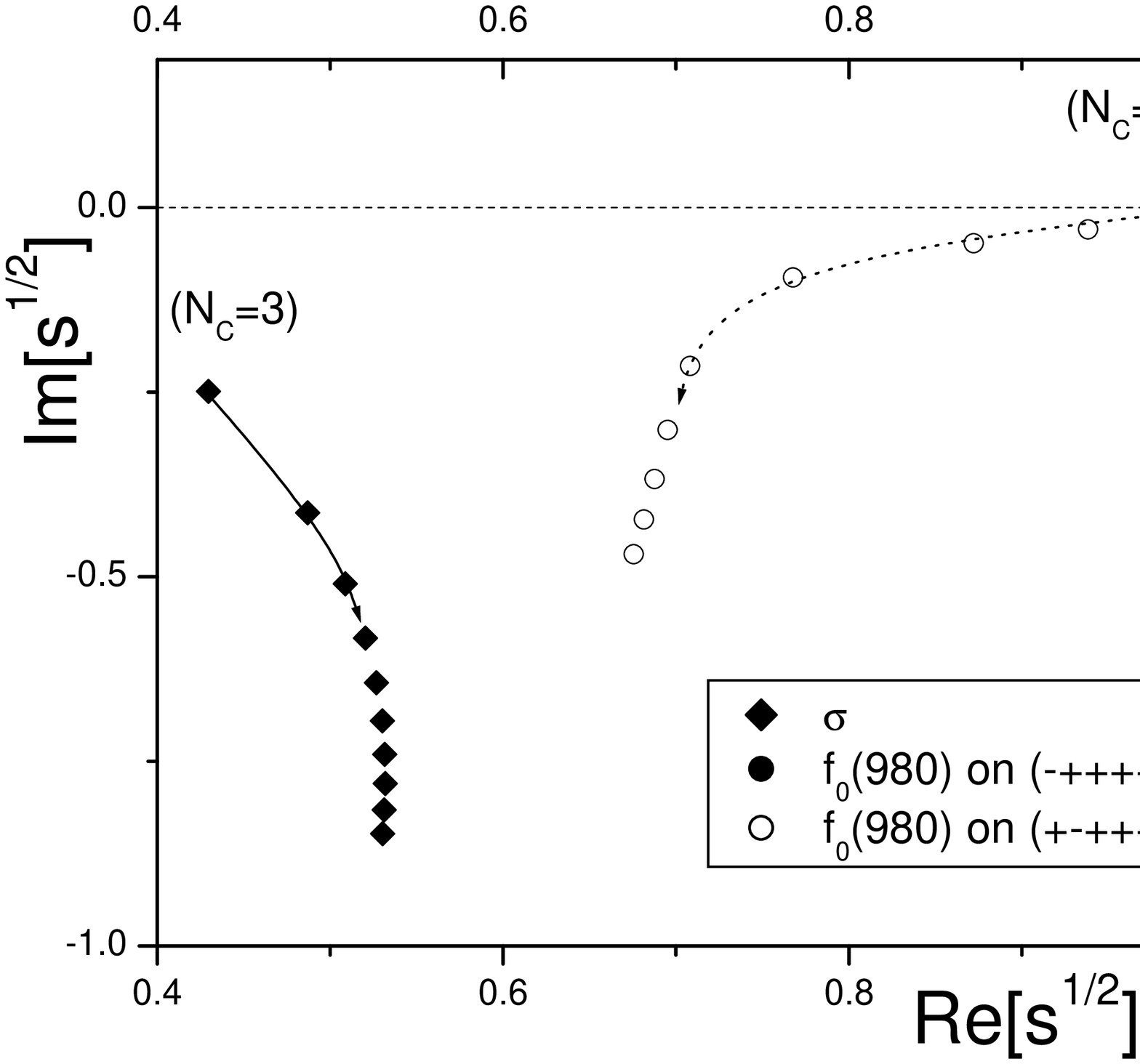}
\caption{\label{kpinc}Left: I=1/2 poles' trajectories; Right: the pole
trajectories of $\sigma$ and $f_0(980)$.}
\end{figure}

The success of describing such a broad range of spectrum and their
decays in a unified and consistent way suggests that it is a
reasonable model to study these resonances and could be used to gain
further insights into their nature.  The large $N_c$ behavior of the
pole trajectories serves to shed light on the origin of these
resonances.  The lowest order of $\alpha$ is $1/\sqrt{N_c}$, $\beta$,
and $\gamma$ by a factor of $1/N_c$ and $1/N_c^2$. The bare mass, the
location of Adler zero, and the form factor are of order 1, while the
mixing angle $\phi$ is of order
$1/N_c$~\cite{'tHooft:1973jz,*Witten:1979kh}.  Whichever of our two
form factors we use, the poles exhibit similar trajectories as $N_c$
increases.  Those for the $I=1/2,\, 0$ resonances are shown in Fig.
\ref{kpinc} as examples. The $\sigma$, $\kappa$, and $a_0(980)$ poles
move farther away from the real axis. In contrast, the $K^*_0(1430)$
and $K^*_0(1950)$ become narrower and move towards the $n\bar{s}$
bound state.  Analogously, the $a_0(1450)$ and $a_0(1830)$ move to the
$(u\bar{u}-d\bar{d})/\sqrt{2}$ bare seeds. The $f_0$ poles other than
$\sigma$ and  $f_0(980)$ move towards either the $n\bar{n}$ or
$s\bar{s}$ bare seeds.  At $N_c=3$, if the coupling to the $\pi\pi$
channel is switched off, the $f_0(980)$ will move down below the
$K\bar{K}$ threshold and form a bound state.  This behavior implies
that the $f_0(980)$ is more like a $K\bar{K}$ molecule state.
However, when $N_c$ increases, it exhibits a peculiar trajectory: it
moves to the real axis rapidly and then crosses the cut onto the
$(+-+++)$ sheet, and then moves away from the real axis as the $\sigma$
pole does, as seen in Fig.\ref{kpinc}. If the coupling to the lowest
thresholds are increased, respectively, by hand when $N_c=3$, the
$\sigma$, $\kappa$, and $a_0(980)$ will move to the real axis and
become virtual bound states different from the seeds either. While the
coupling becomes strong enough, the virtual bound states will move
onto the first sheet and become bound states. It is worth mentioning
that,  in using the IAM to unitarize ChPT~\cite{Pelaez:2003dy},
Pel\'aez has observed similar pole behaviors of $\sigma$ and $\kappa$
in some parameter region. Using the Pad\'e technique to unitarize ChPT
amplitudes, the similar $\sigma$ pole trajectory is also
found~\cite{Sun:2005uk}.  The pole behaviors of $f_0(980)$ and
$a_0(980)$, as we pointed out here for the first time, may explain the
strange behavior of the line shape in large $N_c$ shown in
\cite{Pelaez:2003dy}.

So, the general $N_c$  behavior separates the poles into two types:
$\sigma$, $\kappa$, $a_0(980)$, and $f_0(980)$ are the first type~(or
the unconventional type) of resonances, like bound states of mesons,
which are dynamically generated by the pseudoscalar interactions. This
may be the reason why they could be described by the tetraquark model.
All the other resonances except the glueball candidate, as the second
type~(or the conventional type) of resonances, are directly generated
from $q\bar{q}$ seeds by renormalization effect, which indicates that
they all belong to the same bare $q\bar q$ nonet.  As the interactions
are turned on and different channels are open, the bare seeds are
copied to different Riemann sheets and get renormalized by the hadron
clouds in various ways~\cite{Eden:1964zz}.  Some of them run too far
away from the physical region to be detectable. In this picture, there
is no need to distinguish parts of them to be a nonet.

\section{summary}
In conclusion, this paper demonstrates that the whole low-energy
scalar spectrum below 2.0~GeV, except for a possible glueball
$f_0(1710)$, could be described in one  consistent picture, with the bare
``$q\bar q$ seeds" dressed by the hadron loops. All the resonances are
dynamically generated by the same mechanism, and there is no direct
correspondence between the poles and the original nonet in the
Lagrangian. In a large $N_c$ analysis of this picture, the pole
trajectories exhibit a general behavior which agrees with other
models. In particular, the $\sigma$, $\kappa$, $f_0(980)$, $a_0(980)$
resonances, though running away from the real axis when $N_c$ is
larger, are also generated in this model, which means this large $N_c$
behavior does not conflict with the $q\bar q$ dressed by the hadron loop
picture. They are  produced by large hadron loop effects and this
may also imply their large four-quark components.  Thus, in this paper,
we present that the usual speculation in particle physicist community,
that the lighter scalars behave like the tetraquark states and the heavier scalars do as
the $q\bar q$ states, could be actually realized in such a coherent
picture of improved UQM model.

We also show how the line shape of $a_0(980)$ is possibly generated by
a deep pole, like the $\sigma$ or $\kappa$, encountering the $K\bar{K}$
threshold.  This whole treatment could be extended to other spectra,
e.g., the charmonium states~\cite{Pennington:2007xr}, and provide
theoretical suggestions for further experimental investigation.

\begin{acknowledgments}
We are grateful to Mike~Pennington and Hanqing~Zheng for instructive
discussions and thank  Yanrui
Liu for helpful discussion.   Z.~X. thanks
the Science and Technology Facilities Council in the United Kingdom for
financial support. This work is partly supported by  China Scholarship Council and China
National Natural Science Foundation
under Contracts No.10705009, No.10647113, and No.10875001.
\end{acknowledgments}

\bibliographystyle{apsrev4-1}
\bibliography{scalar2}

\end{document}